\documentstyle[12pt]{article}
\title {\bf  ATOMIC ELECTRON MOTION\\ FOR M\"OLLER
 POLARIMETRY\\ IN A DOUBLE--ARM MODE }
\author{Andrei Afanasev\thanks{On leave from Kharkov Institute of Physics and
Technology, Kharkov 310108, Ukraine}\\ 
{\normalsize\it CEBAF, 12000 Jefferson Ave., 
Newport News, VA 23606, USA} \\
{\normalsize\it and}\\
{\normalsize\it NuHEP Center, Department of Physics, 
Hampton University, Hampton, 
VA 23668, USA}\\ Alexander Glamazdin\thanks{Permanent address: Kharkov 
Institute of Physics and
Technology, Kharkov 310108, Ukraine}\\
 {\normalsize\it CEBAF, 12000 Jefferson Ave., 
Newport News, VA 23606, USA}}
\date{January 31, 1996}
\begin{document}
\maketitle

\begin{abstract}

We analyse an effect of electron Fermi motion at atomic shells 
on the accuracy of 
electron beam polarization measurements with a M\"oller polarimeter operating 
in a double--arm mode. It is demonstrated that the effect can result in either 
{\it increase} or {\it decrease} of the measured polarization depending on the 
detector positions. The effect is simulated for the M\"oller polarimeter to
be installed at CEBAF Hall A.
 \end{abstract}
\vskip 1 in
\centerline{\it To be submitted to Nuclear Instruments and Methods}
\vskip -7 in
\hskip 5 in
{\bf CEBAF--PR--96--003}
\newpage
\section {Introduction}

M\"oller polarimeters are widely used for electron beam 
polarization measurements
in a GeV energy range. High quality of polarization experiments anticipated 
at new--generation CW multi--GeV electron accelerators such as
 CEBAF requires
precise measurements of electron beam parameters. One of these parameters is
the electron beam polarization. It can be measured by a M\"oller polarimeter.

There is a number of systematic corrections which should be accounted for when 
obtaining electron beam polarization from the asymmetry measured with a 
M\"oller polarimeter. They may be related to the knowledge of foil 
magnetization, accidentals, backgrounds, etc. An important systematic 
correction is due to 
electron Fermi--motion at atomic shells \cite{Levchuk}. This correction is in 
principle different from the others listed above because it 
enters on the level 
of elementery $ee$-cross section. By analogy with radiative corrections, this
 correction may be called {\it internal} as opposed to external corrections. 
Therefore, the effect is present in any design of the M\"oller polarimeter, 
operating in 
either single--arm or coincidence modes.

Before the paper \cite{Levchuk} was published, the effect of Fermi--motion 
at atomic shells was belived to negligible. However, the discrepancy between 
polarization  values measured with M\"oller and Compton polarimeters at SLC/SLD
demonstrated that this correction may be large, it was estimated to be 14\%
in this particular case \cite{swartz}.

In this paper, we report on some results obtained while developing a M\"oller 
polarimeter for Hall A at CEBAF. The paper is organized as follows. 
Section 2 describes the formalism for polarized $ee$-scattering and the 
effect of a target electron motion on the spin asymmetry. Section 3 lists 
basic results, and the conclusions are presented in Section 4.
\bigskip  
\section{Formalism }
\bigskip 
M\"oller polarimetry is based on scattering of polarized electron beam on a
polarized electron target. The polarization dependent cross-section for the 
electron-electron scattering is given by \cite{Kresnin}

$$
{{d\sigma^{M\ddot oll}}\over {d\Omega^{*}}} = {{d\sigma_{0}^{M\ddot oll}} 
\over {d\Omega^{*}}} (1+ \sum_{i,j}P^{b}_{i} A_{ij}P^{t}_{j}),\eqno(1)
$$

$$
(i,j = x,y,z),
$$
where $P^{b}_{i}$ ($P^{t}_{j}$) are components of the beam (target) 
polarization, $A_{ij}$ are the asymmetry parameters, and  ${{d\sigma}
_{0}^{M\ddot oll}}/ {d\Omega^{*}}$ is the cross--section for the unpolarized 
particles. Here we use the coordinate system with $z$--axis the  
electron beam direction and $x$--($y$-) axis coplanar (normal) to the reaction
 plane.

Using one--photon exchange and the ultrarelativistic limit 
for the unpolarized cross--section and the nine asymmetry parameters one has

$$
{{d\sigma_{0}^{M\ddot oll}}\over {d\Omega^{*}}} = {{\alpha^{2}}\over{4m^{2}}}
\gamma^{-2}{{(4 - \sin^{2}\Theta^{*})^{2}}\over{\sin^{4} \Theta^{*}}},\eqno(2)
$$

$$
{A_{zz}} = -{{(7+\cos^{2}\Theta^{*})\sin^{2}\Theta^{*}}\over{(3+\cos^{2}
\Theta^{*})^{2}}},\eqno(3)
$$

$$
{-A_{xx}} = {A_{yy}} = {{\sin^{4}\Theta^{*}}\over{(3+\cos^{2}\Theta
^{*})}^{2}},\eqno(4)
$$

$$
{A_{zx}} = {A_{xz}} = -{{2\sin^{3}\Theta^{*}\cos\Theta^{*}}\over
{\gamma(3+\cos^{2}\Theta^{*})^{2}}},\eqno(5) 
$$

$$
{A_{xy} = A_{yx} = A_{zy} = A_{yz} = 0},\eqno (6)
$$

$$
{\gamma = \sqrt{{(E_{0}+m)}\over{2m}}},
$$
where $\alpha$ is the fine--structure constant, $\Theta^{*}$ is the 
c.m.s. scattering angle, $m$ is the electron mass, and $E_{0}$ is the  
energy of the incident electron in the laboratory  system.

It is seen that at $\Theta^{*} = 90^{0}$ the asymmetry parameters $A_{xx}$, 
$A_{yy}$ and $A_{zz}$ are maximal

$$
A_{zz} = -{7\over 9},\ \ \ A_{xx} = -{1\over 9},\ \ \ A_{yy} =
{1\over 9}, \eqno(7)
$$
and the asymmetries $A_{xz}$ and $A_{zx}$ are small within the whole angular 
acceptance and vanish at $90^0$.

In experiment, the polarized electron beam is incident on a magnetized 
ferromagnetic foil. The observed symmetry of M\"oller scattering from atomic 
electrons,

$$
A={N_{\uparrow \uparrow} - N_{\uparrow \downarrow}\over N_{\uparrow
 \uparrow} + N_{\uparrow \downarrow}},\eqno(8)
$$
gives the desirable polarization of the electron beam provided the target 
polarization is known.

Let us consider M\"oller scattering on a moving target electron from 
a particular atomic shell $n$. In the laboratory frame, cosine of the 
scattering angle is given by

$$
\cos\Theta = \cos\Theta_{0} + {(\Delta\cos\Theta)}_{F},\eqno(9)
$$
where $\Theta$ $(\Theta_{0})$ is a lab. scattering with (without) 
target electron motion, and the $(\Delta\cos\Theta)_{F}$ is a
correction due to Fermi--motion of the target (denoted by $P_{F}$), and 
neglecting higher--order terms in expansion over $E^{-1}_{0}$, we obtain

$$
{(\Delta\cos\Theta)}_{F} = \cos\Theta_{12}{P_{F}^{n} \over m}
{(1 - \cos\Theta_{0})},\eqno(10)
$$
where $\Theta_{12}$ is an angle between the momenta of the beam and the target 
electron. At $\Theta^*=90^{0}$, Eq.(10) becomes

$$
{(\Delta\cos\Theta )_{F} = \cos\Theta_{12}{P_{F}^{n}\over E_{0}}},\eqno(11)
$$

this result was reported earlier \cite{afanas}. Eq.(11) can also 
be rewritten in the form        

$$
\Theta=\Theta_{0}\sqrt {1 - {P_{F}^{n} \over m}\cos \Theta_{12}},\eqno(12)
$$

In the leading order of $P_{F}$/$m$ expansion, it reproduces the 
original L. Levchuk's result (Eqs. 9-10 of Ref.\cite{Levchuk}). However, for 
the case of large acceptances it is more consistent to use a general result
Eqs. (9)-(10) herein, because angles far from $\Theta^{*}=90^{0}$ are 
involved. 
Let us summarize the obtained results. Atomic electron motion does not 
affect the values of cross section and scattered electrons energies but it 
changes the angle of M\"oller scattering. A dominant effect comes from target 
electron motion parallel (antiparallel) to the direction of the incident beam; 
an effect from transverse target motion is supressed by an extra factor of 
$\sqrt{2m/E_{0}}$. Using the changed M\"oller kinematics described by 
Eqs.(9)--(10), one should take a sum over all atomic shells and integrate over 
$\Theta_{12}$ and $P_{F}$ using a proper atomic wave function. To estimate the
 effect, we assume what $P_{F}^{n}=\sqrt{2m\varepsilon_{B}^{n}}$,  
$\varepsilon_{B}^{n}$ being an electron binding energy at the atomic shell 
$n$.

Electrons in the atom of iron have the binding energies \cite{aip} listed in 
Table 1.
\bigskip

Table 1. Electron binding energies for $^{26}Fe$ (atomic moment=2.22$\mu_{B}$)
\medskip

\centerline{\begin {tabular}{lcccccccc}
Shell &$K(1s)$ & $L_I(2s)$ & $L_{II}(2p_{1/2})$ & $L_{III}(2p_{3/2})$ & 
$M_I(3s)$ & $M_{II,III}(3p)$ & 
$M_{IV, V}(3d)+N_I(4s)$ \\[0.2cm]
$\varepsilon_{B}^{n}$, eV& 7112& 846.1& 721.1& 708.1 & 92.9&
54.0 & 3.6$\pm$0.9 \\[0.2cm]
Number of&2&2&3&3&2&6&$6(3d)+2(4s)$\\
electrons& & & & & & &\\
\end{tabular}}
\vskip 0.25 in

Thus, scattering from $K$- and $L$-shells smears the M\"oller scattering 
angle by $\Delta\Theta / \Theta\simeq\pm10\%$ and $\pm3\%$ respectively, 
around 
$\Theta^{*}=90^{0}$. Only the electrons on the incomplete $M$-shell are 
polarized. (Iron needs 4 more electrons on $3d$-shell to complete it. 
Overlapping $3d$ and $4s$ levels in metal yield the observable hyromagnetic 
ratio of 2.22$\mu_{B}$). The M\"oller asymmetry for electron-atom scattering 
may be presented as 

$$
A={{\Sigma (\sigma_{n}(\uparrow\uparrow )-\sigma_{n}(\uparrow\downarrow ))} 
\over {\Sigma(\sigma_{n}(\uparrow\uparrow ) + \sigma_{n}(\uparrow\downarrow 
))}},\eqno(13) 
$$
where the sum is taken over all $n$ atomic shells, and $\sigma_{n}{(
\uparrow\uparrow )}$ or $\sigma_{n}{(\uparrow\downarrow )}$ 
correspond to the M\"oller cross section with spins parallel or antiparallel,
 respectively. Only polarized (loosely bound) electrons from the $M$--shell 
contribute to the numerator in Eq.(13) r.h.s., whereas all electrons including 
strongly bound $K$--, $L$--shell ones, contribute to the denominator of 
Eq.(13) r.h.s.
 Therefore, kinematic smearing due to Fermi--motion affects the denominator, 
rather than the numerator for the asymmetry expression Eq.(13). 

For a standard geometry of M\"oller polarimeters,when the detector(s) is(are) 
centred around $\Theta_{cm}=90^{0}$, it may result in missing electrons 
scattered from $K$--, $L$--shells yielding a higher theoretical estimate for 
the asymmetry $A$. It was the basic conclusion of \cite{Levchuk} confirmed 
later in SLAC/SLC measurements \cite{swartz}. However, in contrast to 
original predictions \cite{Levchuk}, this effect does not exceed 2\% for
 single--arm M\"oller measurements at MIT  \cite{arrington}.
For the double--arm M\"oller polarimeter
at MIT \cite{beard} a preliminary estimate done by one of us  
\cite{afanas} predicts the correction to the M\"oller asymmetry
to be around 12\% , but this estimate
may change if the magnetic field of the polarimeter quadrupole and 
boundary conditions are carefully included into the calculations.

We found the effect to be strongly dependent on positioning the detectors 
in a double--arm M\"oller polarimeter.
  
\section{Dependence on the detector positions }
 
The 'smearing' of the kinematics due to target electron motion changes 
asymmetry of the M\"oller scattering. This effect should be carefully 
calculated for any specific experimental set--up, since the magnitude of the 
effect depends on target thickness (via multiple scattering), polarimeter 
acceptances, magnetic optics, electron beam parameters, $etc.$

In our study, we have found a new effect due to Fermi--motion of atomic 
electrons. The effect is a dependence of the measured M\"oller asymmetry 
on the relative position of detectors in a double--arm mode. This effect 
can be understood from Figs. 1 and 2. Fig. 1 demonstrates the ratio of M\"oller
asymmetry neglecting the effect of electron Fermi--motion to the same 
quantity but with the Fermi--motion included in the calculation, as a function 
of the detectors displacement from the symmetric around $\Theta^{*}=90^{0}$ 
positions. Note that for the chosen electron energies and target thickness, 
the angular smearing due to Fermi--motion for $K$--electrons is an order of 
magnitude larger than due to multiple scattering in the target.
Different displacements of the detectors in a double--arm mode 
result in a different effect due to Fermi motion on the M\"oller asymmetry. 
The correction is {\it positive} and reaches its maximum value for the case of 
symmetric position of the detectors, centered at the angle corresponding to 
$90^{0}$--scattering angle in c.m.s.(denoted A in Fig. 2.)
If the detectors are 
moved simultaneously toward larger (position B) or smaller (position C) angles,
the Fermi--motion correction to the M\"oller asymmetry becomes {\it negative}. 
In absence of multiple scattering and the displacements lager than B or smaller
than C ($e.g.$, position D), the Fermi--motion 
correction to the asymmetry would 
be exactly --100\%. It means that we can observe a {\it zero} 
M\"oller asymmetry 
scattering polarized electron beam on atomic electrons with {\it nonzero} net 
polarization! The reason is that for this geometry of the experiment, we 
detect only the electrons scattered on unpolarized atomic shells. This 
polarization asymmetry of the detected electrons if 
$|\Delta\Theta_d/\Delta\Theta_{acc}|\geq 0.5$ is completly due to multiple 
scattering of electrons 
in the target. It may provide a direct measure of multiple scattering effect
 ($i.e.$, target thickness). Further displacing the detectors, the multiple 
scattering effect dies off exponentially, and the asymmetry of the electrons 
detected in coincidence approaches zero, but we may still observe a 
considerable amount of M\"oller electrons in coincidence coming from the 
tails of momentum distributions for atomic electrons.

The calculations in Fig. 1 were done for illustration, treating the 
polarimeter schematically as a target + a pair of detectors, with no magnets 
involved in the system. The acceptance (normalized to unity), 
the Fermi--motion 
smearing and multiple scattering angles were the same as for the realistic 
case described below.

Simulation of the Fermi--motion effect was done for the M\"oller polarimeter 
of CEBAF Hall A. A detailed description of this polarimeter is given in Ref. 
\cite{Design}. The polarimeter is designed for coincidence mode operation. 
The magnetic system includes two quadrupoles and a dipole. The simulation was 
done by RAYTRACE combined with a Monte--Carlo code for simulating M\"oller 
and multiple scattering. The results of the simulation are presented in 
Figs. 3, 4 for the electron beam energy $E_0=$ 0.8 GeV and target thickness
 $= 17.6 \mu$m. The distribution of M\"oller electrons in the detector 
plane is demonstrated in Fig.3 for one of the detectors. For the other 
detector, the distribution is symmetric. The axis $X$ is perpendicular to  the
reaction plane, and $Y$--axis indicates displacement with respect to the beam
axis. The square area dashed with lines having a positive slope demonstrates 
the detector 
centered at $Y_0$ corresponding to $\Theta^*=90^0$ ($\Theta_A= 35.7$ mrad).
It provides the angular acceptance $\Delta\Theta/\Theta$=10.5\%. The
square dashed with lines having a negative slope demonstrates the 
detector displaced along the $Y$--axis
by distance $d$. The second--arm detector is displaced symmetrically with
respect to the beam axis. The effect of this displacement is shown in Fig.4
and appears to be in qualitative agreement with the results of Fig.1
obtained for a simplified model of the polarimeter. As can be seen from Fig.4,
the Fermi--motion effect is positive and maximal, reaching $\approx$10\%,
for small displacement $d$, wereas for large values of $d$, the effect is
negative and may reach the magnitude of $-100$\% completely eliminating 
the M\"oller asymmetry. The plot in Fig.4 is asymmetric with respect to
$d$=0 due to dipole dispersion (it would become symmetric if plotted $vs.$
angular shifts). The plateau (instead of a maximum like in Fig.1) is caused
by additional boundary conditions set by the polarimeter magnetic system
acting like an effective collimator.

It should be noted that the larger is the angular acceptance, the less
Fermi motion affects the measured M\"oller asymmetry. We choose the beam energy
$E_0$= 0.8 GeV for illustration because for higher energies, the angular 
acceptance of the polarimeter becomes large (25--42\% for $E_0$=1.6--6.0 ~GeV)
and the maximum positive Fermi--motion effect is small, not exceeding 3\%.

\bigskip  
\section{Summary}
\bigskip 
We studied an effect of atomic electron Fermi--motion for a double--arm 
M\"oller polarimetry. We demonstrate that this effect may be either positive 
or negative depending on positioning of the detectors. If the detectors are
centered at $90^0$--
scattering angle (in c.m.s.), the correction has a maximum positive value. 
For detectors shifted simultaneously toward larger/smaller angles, the 
effect becomes negative and may reach --100\% completely eliminating the
observed spin asymmetry. On the other hand, for a single--arm measurement 
the Fermi--motion correction remains positive despite the shift in 
the detector position.

Magnetic fields may essentially change electron kinematics, therefore, it is 
necessary to do detailed simulation of the polarimeter optical system in order 
to consistently calculate the electron Fermi--motion effect.

\bigskip\bigskip
{\large\bf Acknowledgements}
\bigskip

\noindent This work was supported by the US Department of Energy under
contract DE--AC05--84ER40150.

 A.A. would like to
acknowledge useful discussions with his colleagues at CEBAF, Hampton 
University and KhPTI.
A.G. would like to thank V.G.~Gorbenko for collaboration in
writing the Monte--Carlo simulation code for CEBAF Hall A M\"oller polarimeter.

\vspace{2cm}
{\large\bf Figure Captions}
\bigskip

{\bf Figure 1.} Fermi--motion 
correction to the M\"oller asymmetry for $^{26}Fe$
target as a function of detectors displacement, $\Delta\Theta_d$ being the
displacement angle with respect to $\Theta^*$=
90$^0$, $\Delta\Theta_{acc}$ being the angular acceptance, and $A(A_0)$ being 
the asymmetry with (without) the Fermi--motion 
correction. 
\bigskip

{\bf Figure 2}. Schematic positions of the detectors. The angle $\Theta_A$ 
corresponds to $\Theta^*$= 90$^0$, and the positions A, B, C, and D correspond
to $\Delta\Theta_d/\Delta\Theta_{acc}$= 0, 0.5,--0.5, $>$0.5, respectively,
 in Fig.1.
\bigskip

{\bf Figure 3}. Simulated distribution of M\"oller events in the detector 
plane of CEBAF
Hall A M\"oller polarimeter. Dashed square areas show different positions
of the detectors. The notations are explained in the text.
\bigskip

{\bf Figure 4}. Simulated Fermi--motion effect for 
CEBAF Hall A M\"oller polarimeter
as a function of the detectors displacement $d$, as shown in Fig.3.

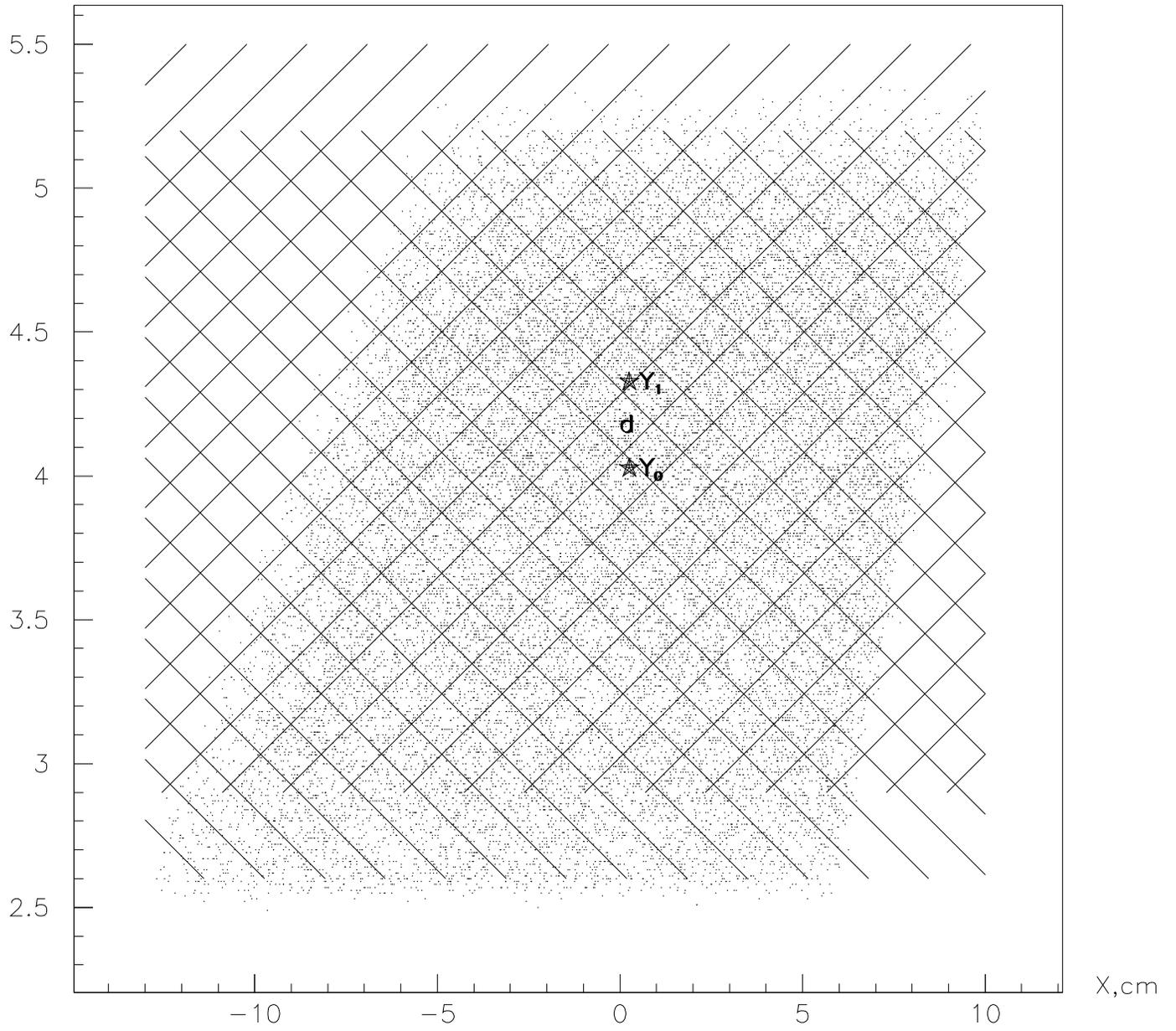
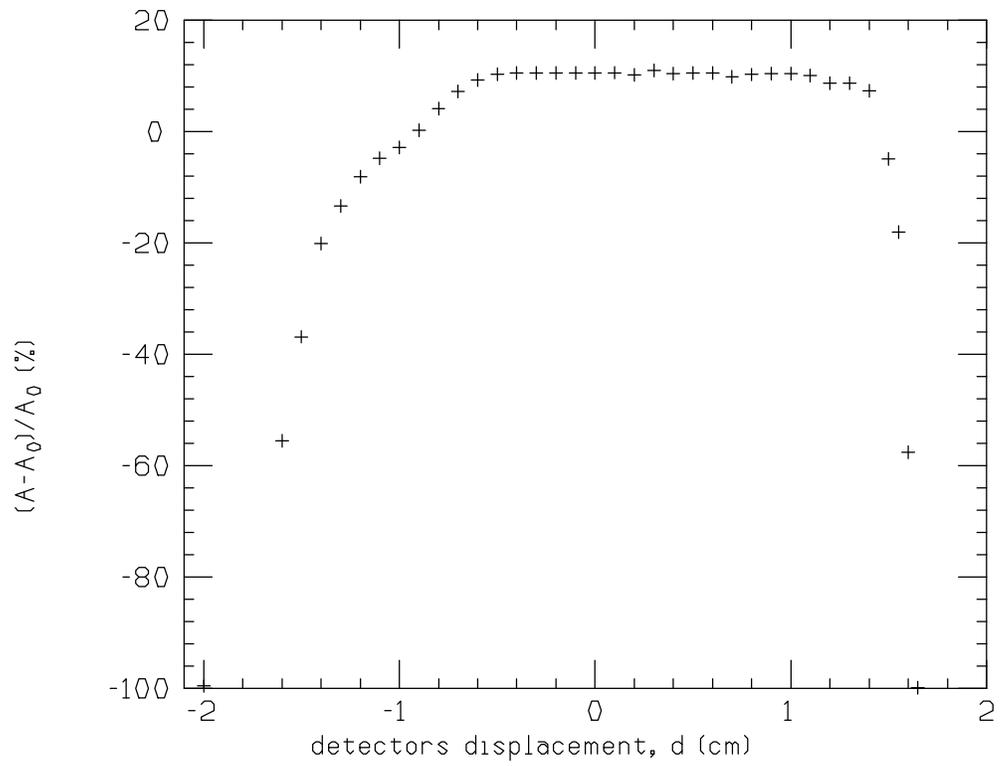
\end{document}